\pdfoutput=1

\documentclass[11pt]{article}

\usepackage[preprint]{acl}
\usepackage{amsmath}
\usepackage{amssymb}
\usepackage{graphicx}
\usepackage{subcaption}

\usepackage{times}
\usepackage{latexsym}

\usepackage[T1]{fontenc}

\usepackage[utf8]{inputenc}

\usepackage{microtype}

\usepackage{inconsolata}

\title{The Community Archive: An Ethically Sourced X (Twitter) Dataset}

\author{
  Brandon Duderstadt\\
  Calcifer Computing\\
  \texttt{brandon@calcifercomputing.com}
  \And
  Omar Shehata\\
  Open Memetics Institute\\
  \texttt{omar.sameh.shehata@gmail.com}
  \AND
  Ivan Vendrov\\
  Supercoordination\\
  \texttt{ivendrov@gmail.com}
  \And
  Francisco Carvalho\\
  Epistemic Garden\\
  \texttt{xiq@epistemic.garden}
}

\begin{document}
\maketitle

\begin{abstract}
We present The Community Archive, a novel dataset of over 5 million posts from the social media site X (previously known as Twitter).
In contrast with other large scale social media datasets, which are generally assembled without user consent via web scrapers, The Community Archive uses an opt-in model to obtain author consent for every post in the dataset.
This enables us to release both the metadata and the content of users' posts at a previously unprecedented scale. 
Our dataset supports work in a wide variety of domains, including social network analysis, narrative tracking, and diachronic language modeling. 
We demonstrate the utility of our dataset by conducting a sample analysis of content-based homophily in our network.
\end{abstract}

\section{Introduction}
Social media platforms have become deeply embedded in daily life, serving as key arenas for communication, information dissemination, and narrative formation.
As the influence of these platforms continues to grow, so too does the need for open, accessible datasets that allow researchers to rigorously study their social, linguistic, and network dynamics.
Despite the scientific importance of such datasets, the field faces a critical bottleneck; legal and ethical barriers severely restrict the distribution of social media data at scale.
This problem is becoming more acute as social media platforms increasingly close their ecosystems to external research access.
Further, existing large-scale social media datasets are often assembled through web scraping without explicit user consent.
This creates significant legal and ethical concerns that limit widespread dataset adoption.

In this paper, we introduce The Community Archive, a novel dataset of over 5 million posts collected from the social media site X (formerly Twitter) through a fully opt-in process. 
Every post included in the dataset was contributed with explicit author consent, allowing us to release both post content and metadata at an unprecedented scale.
This opt-in approach directly addresses the legal and ethical barriers that have historically limited dataset availability.

Overall, we view the Community Archive as an X (Twitter) analogue to the popular "Zachary's Karate Club" \cite{zachary1977information} dataset.
Notably, we do not claim that the posts in the Community Archive are representative of the discourse on X (Twitter) as a whole.
Rather, we view The Community Archive as a resource that supports research in a broad range of domains, including social network analysis, narrative tracking, and diachronic language modeling.
We demonstrate our dataset's utility by conducting an analysis of content-based homophily, or the tendency of users who talk about similar content to be connected, in our network.

\begin{table*}[h]
\centering
\caption{A comparison of several widely used X (Twitter) datasets from the literature. Most datasets either lack the text of source posts or lack corresponding metadata like number of likes and follower-following relationships. Existing datasets that contain both post text and metadata were collected without user consent, limiting their applicability. The Community Archive is the first large X (Twitter) dataset that contains both post text and metadata from consenting users.}
\label{tab:dataset_comparison}
\vspace{0.5cm}
\begin{tabular}{|l|l|l|l|l|}
\hline
\textbf{Dataset} & \textbf{Size} & \textbf{Has Post Text} & \textbf{Has Metadata} & \textbf{Consensual} \\ \hline
\cite{abilov2021voterfraud2020multimodaldatasetelection} & 7.6 Million   & No & \textbf{Yes}               & No \\ \hline
\cite{Chen_2020} & 123 Million   & No & \textbf{Yes}               & No \\ \hline
\cite{Lamsal_2020} & 200 Million   & No & \textbf{Yes}               & No \\ \hline
\cite{zubiaga2016rumours} & 4,842 & \textbf{Yes} & \textbf{Yes}               & No \\ \hline
\cite{derczynski-etal-2016-broad} & 10,000 & \textbf{Yes} & No & No \\ \hline
\cite{barbieri2020tweetevalunifiedbenchmarkcomparative} & 200,785 & \textbf{Yes} & No & No \\ \hline
\cite{muennighoff2023mtebmassivetextembedding} & 79,543 & \textbf{Yes} & No & No \\ \hline
\cite{archiveteam_twitterstream} & 1 Billion + & \textbf{Yes} & \textbf{Yes} & No \\ \hline
\cite{feng2023twibot22graphbasedtwitterbot} & 88,217,457 & \textbf{Yes} & \textbf{Yes} & No \\ \hline
The Community Archive \textbf{(Ours)}& 5,374,594 & \textbf{Yes} & \textbf{Yes} & \textbf{Yes} \\ \hline
\end{tabular}

\end{table*}

\section{Related Work}
The critical role of datasets in enabling the study of social media properties has been recognized for many years.
In response, a number of datasets have been assembled to support this research.

Several datasets focus on releasing large collections of tweetIDs and metadata, but not the content of the posts themselves. 
The VoterFraud2020 dataset \citep{abilov2021voterfraud2020multimodaldatasetelection} collects 7.6 million tweets and 25.6 million retweets related to voter fraud, releasing metadata such as perceptual image hashes, estimated author clusters, and suspension statuses, but notably omitting the original tweet text.
Similarly, \cite{Chen_2020} release 123 million tweet IDs associated with tweets about the COVID-19 pandemic, and \cite{Lamsal_2020} release tweet IDs and sentiment scores for over 400 million COVID-19-related tweets.
Because these datasets only distribute tweet IDs and not post content, they shift the onus of copyright compliance and license acquisition onto the dataset consumer, who must rehydrate the data themselves under X (Twitter)'s terms.

Some datasets do publish the content of posts, but they are often much smaller and more specialized than general X (Twitter) dumps.
The PHEME dataset \citep{zubiaga2016rumours} compiles 4,842 tweets related to rumors about nine newsworthy events, with manual annotations of rumor factuality.
The Broad Twitter Corpus \citep{derczynski-etal-2016-broad} releases the text of 10,000 tweets sampled from several geographic areas, with a focus on named entity recognition.
The TweetEval benchmark \citep{barbieri2020tweetevalunifiedbenchmarkcomparative} collects tweets annotated for tasks like emoji prediction, sentiment analysis, and hate speech detection, but does not provide post metadata.
The Massive Text Embedding Benchmark (MTEB) \cite{muennighoff2023mtebmassivetextembedding} include tweet-based subtasks that provide post text but again omit post metadata.
Overall, the limited size and scope of these corpora restrict their utility for broad social network analysis.

Some resources attempt to offer broader coverage but face legal or ethical limitations.
The Internet Archive Twitter Stream \citep{archiveteam_twitterstream} hosts a very large dataset containing tweet texts and metadata spanning over a decade.
Similarly, the Twibot-22 dataset 
\citep{feng2023twibot22graphbasedtwitterbot} distributes post data alongside richly annotated metadata for the purpose of bot detection.
Unfortunately, both of these datasets were collected without explicit user consent, which restricts their widespread adoption.

Table \ref{tab:dataset_comparison} summarizes the contributions of the aforementioned datasets.
Despite the range of available datasets, there remains a need for a broad, ethically sourced dataset of Twitter posts that includes both post content and metadata.
Such a dataset would enable large-scale, responsible research into both the linguistic and network properties of social media while avoiding the legal and ethical challenges posed by unauthorized data collection.
To remedy this need, we introduce The Community Archive.

\begin{figure*}[t]
\centering
\begin{minipage}{\textwidth}
    \centering
    \begin{subfigure}{0.5\textwidth}
        \centering
        \includegraphics[width=\textwidth]{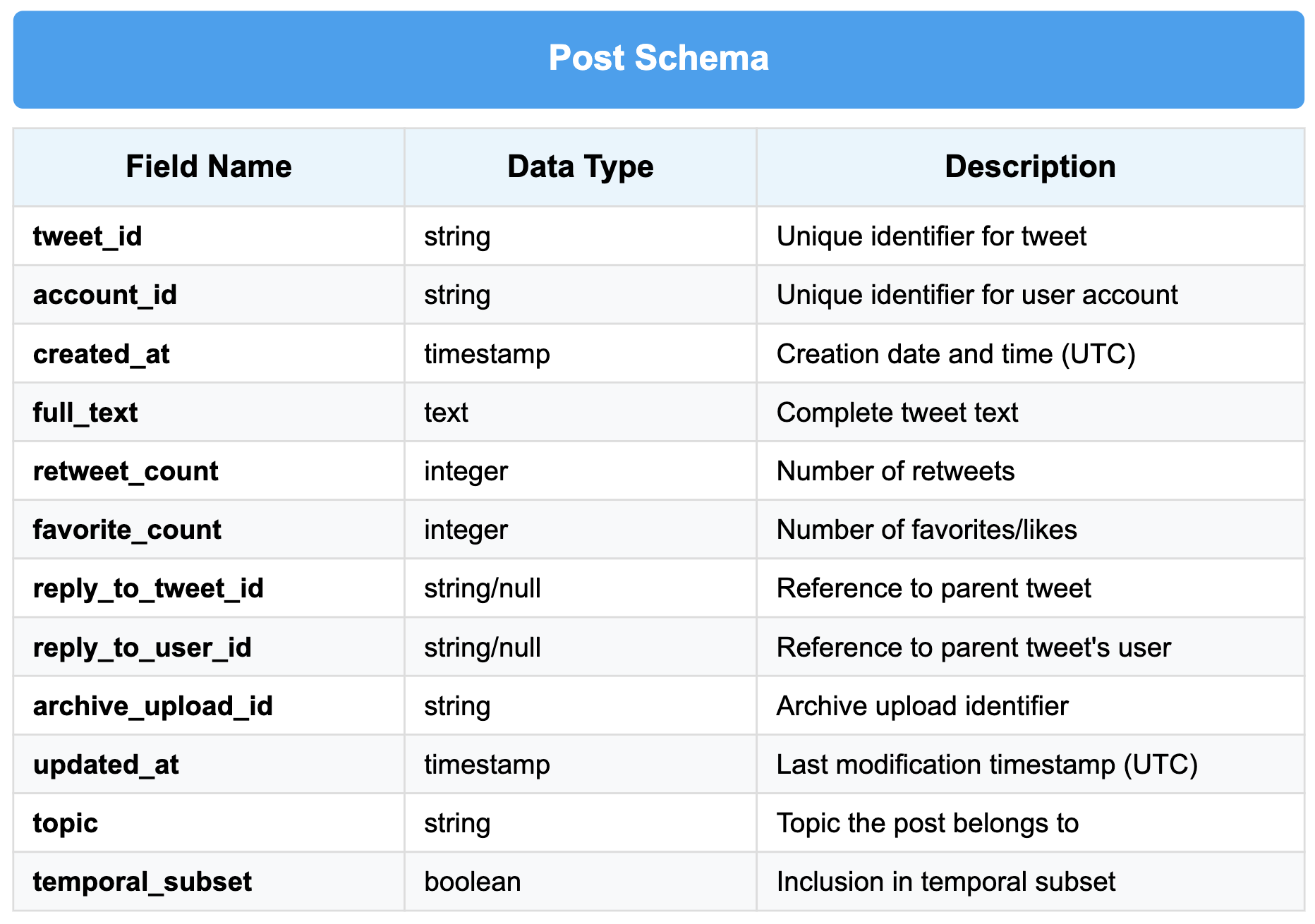}
        \vspace{0.69cm}
        \label{fig:subfig1}
    \end{subfigure}%
    \begin{subfigure}{0.5\textwidth}
        \centering
        \includegraphics[width=\textwidth]{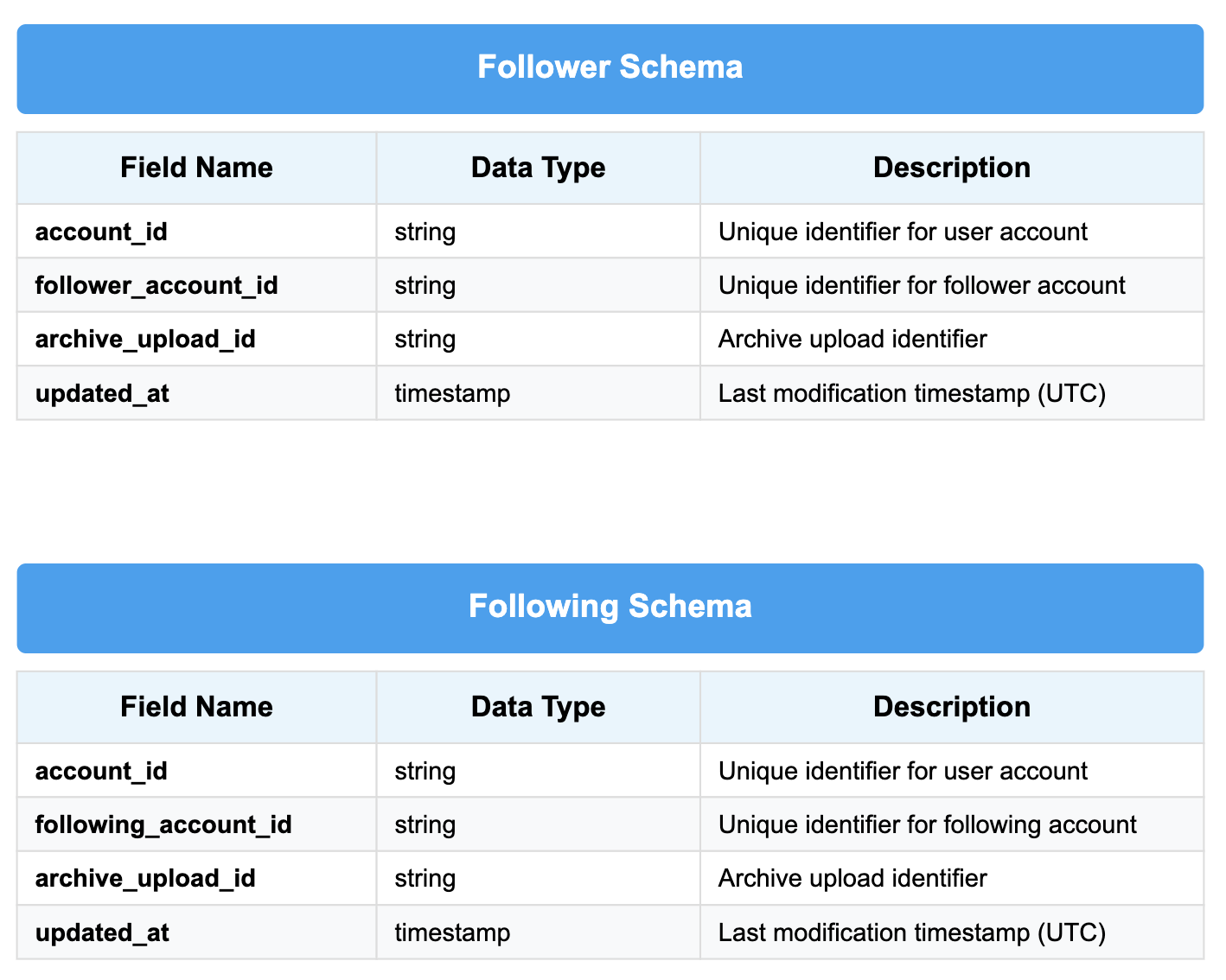}
        \label{fig:subfig2}
    \end{subfigure}
\end{minipage}
\caption{The Community Archive database schema. The database contains tables that include full post data as well as follower/following relationships in the social graph.}
\label{fig:schema}
\end{figure*}

\section{The Community Archive}
\subsection{Collection Methodology}
The Community Archive is a collection of over 5 million posts from 244 unique accounts who have opted in to having their X (Twitter) openly distributed. 
Users donate both the content of their posts as well as post and account metadata to The Community Archive, enabling us to compile a rich dataset containing both social graph and post content data.
All collected data is released under an MIT License.

To contribute to The Community Archive, users first export their individual archive from the X (Twitter) platform.
After a few days, they receive an email from the X (Twitter) team containing their archive as a zip file.
Users then upload this zip file using the application at community-archive.org
Finally, user archives are compiled into a unified database using the schema depicted in Figure \ref{fig:schema}.
Users that wish to revoke their data donation can do so at any time using the delete feature in the online interface (see Appendix Figure \ref{fig:delete}).
Note that the topic field of each tweet is inferred using a process we outline in Section \ref{sec:stats}.

As a result of the asynchronous nature of our opt-in process, different users' archives have been exported from X (Twitter) at different times.
This nuance slightly complicates temporal corpus analysis; for example, the decay of a particular topic in the dataset could be a result of either a decrease in the topic's popularity or the sampling bias in our dataset (e.g. consider the case where an author posted about a topic after submitting their archive).

To correct for this, we create a subset of the data we call the temporal subset, which is appropriately filtered for temporal analysis.
The temporal subset contains posts which were posted before June 1 2024 by authors who created their archives after June 1 2024.
This ensures that, if an author in the temporal subset lacks a post about a particular topic, it is because they did not post about it at all in that time period.
The temporal subset remains quite large, retaining over 86\% of the dataset.

We make two anonymization modifications to the dataset.
Since X (Twitter)'s terms of service prohibit the bulk distribution of internal ids, we hash all tweet and account ids before distribution.
This ensures that we retain the follower-following and reply structure of the posts without violating the X (Twitter) terms of service.
We also replace account names with anonymized user tokens.
This is particularly important for maintaining the privacy of of users who are mentioned in the tweets of opt-in users, but have not opted-in themselves.

\begin{figure*}[t]
\centering
\begin{minipage}{\textwidth}
    \centering
    \begin{subfigure}{0.49\textwidth}
        \centering
        \includegraphics[width=\textwidth]{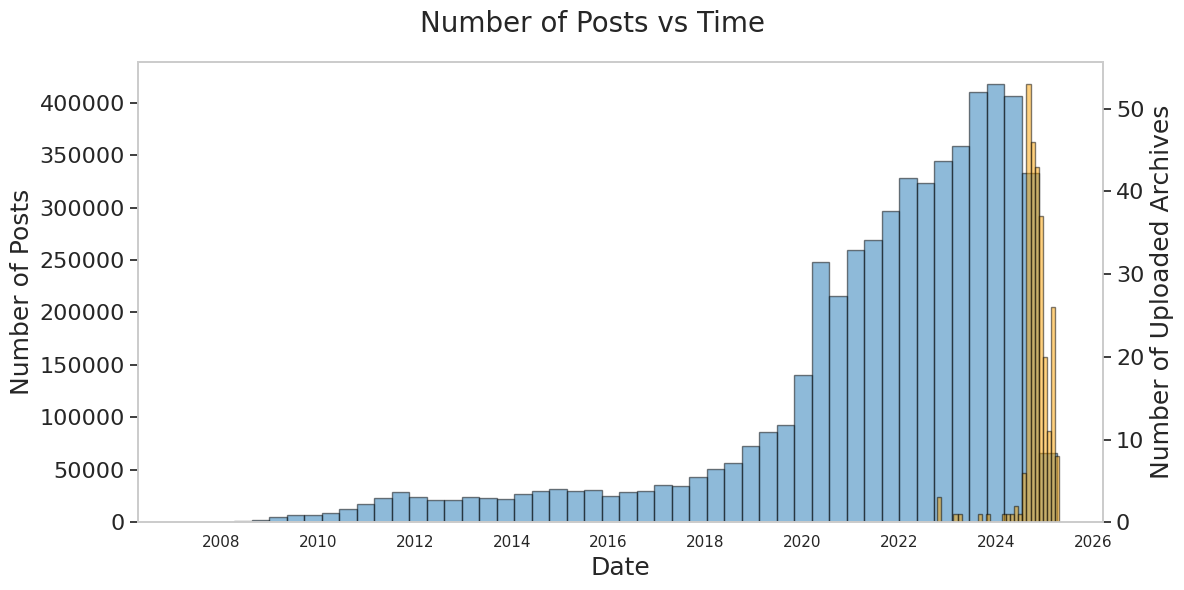}
    \end{subfigure}%
    \hspace{0.01\textwidth}
    \begin{subfigure}{0.49\textwidth}
        \centering
        \includegraphics[width=\textwidth]{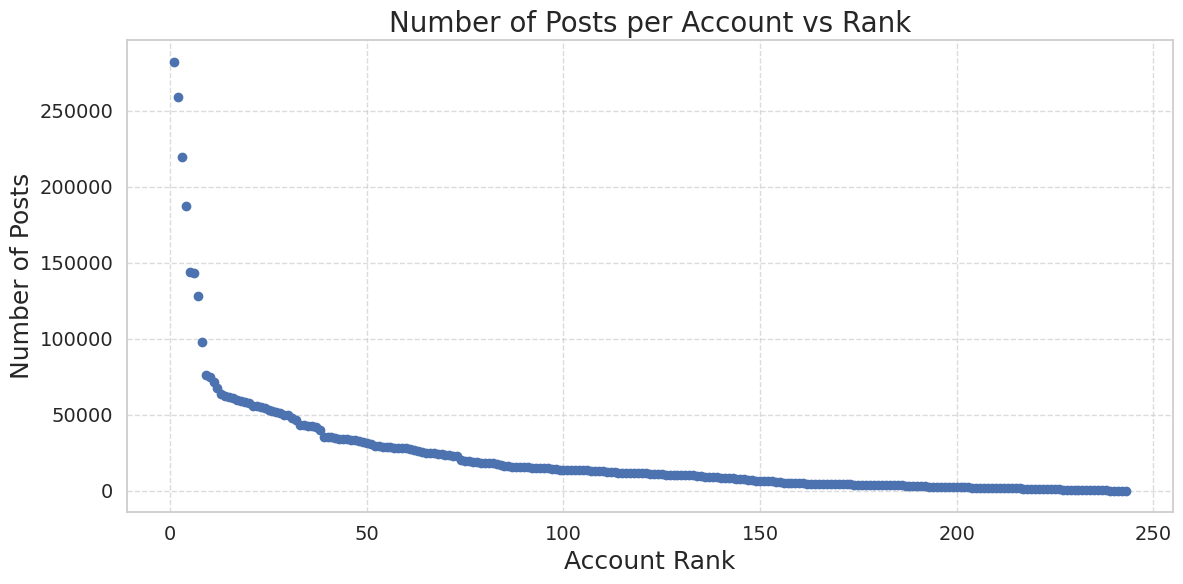}
    \end{subfigure}
\end{minipage}
\caption{\textbf{Left:} The number of posts (blue) and contributed archives (orange) vs time in The Community Archive. 71.1\% of posts occur between January 1 2020 and June 1 2024. Post availability starts to drop as archive upload activity drops. \textbf{Right:} The number of posts for each account vs the account's rank (e.g. the top posting account is rank 1). The top 20\% of authors account for 64.21\% of the posts. The top 5 accounts by number of posts are annotated.}
\label{fig:twitstats}
\end{figure*}

\begin{figure*}[t]
\centering
\includegraphics[width=\textwidth]{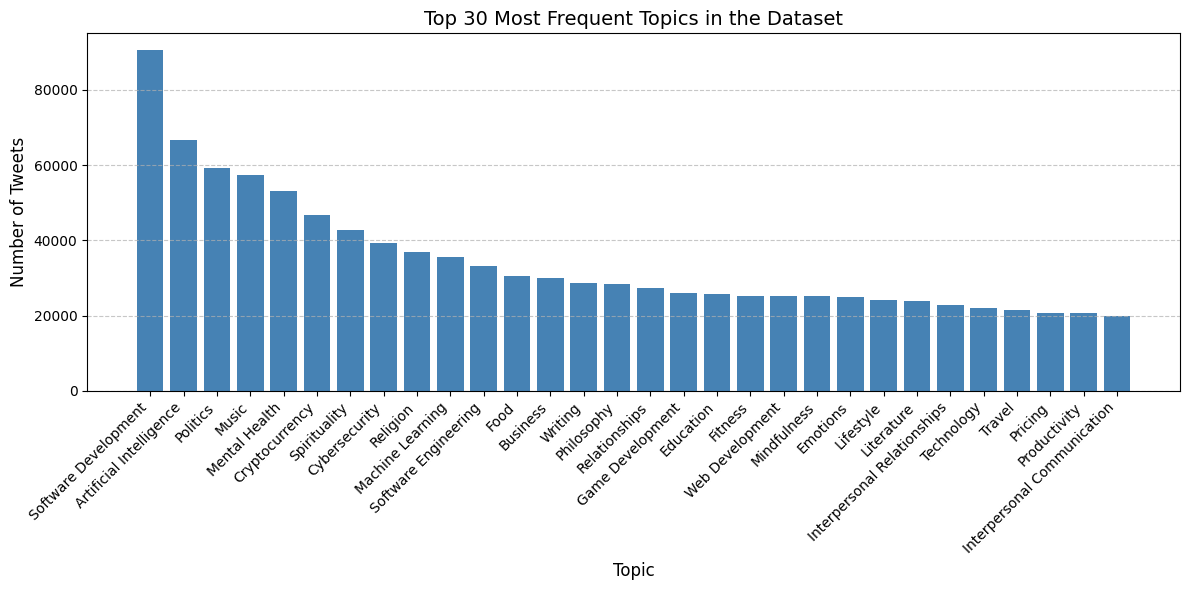}
\caption{The number of posts in each of the 30 most frequent topics in The Community Archive.}
\label{fig:topics}
\end{figure*}

\begin{figure*}[t]
\centering
\includegraphics[width=0.8\textwidth]{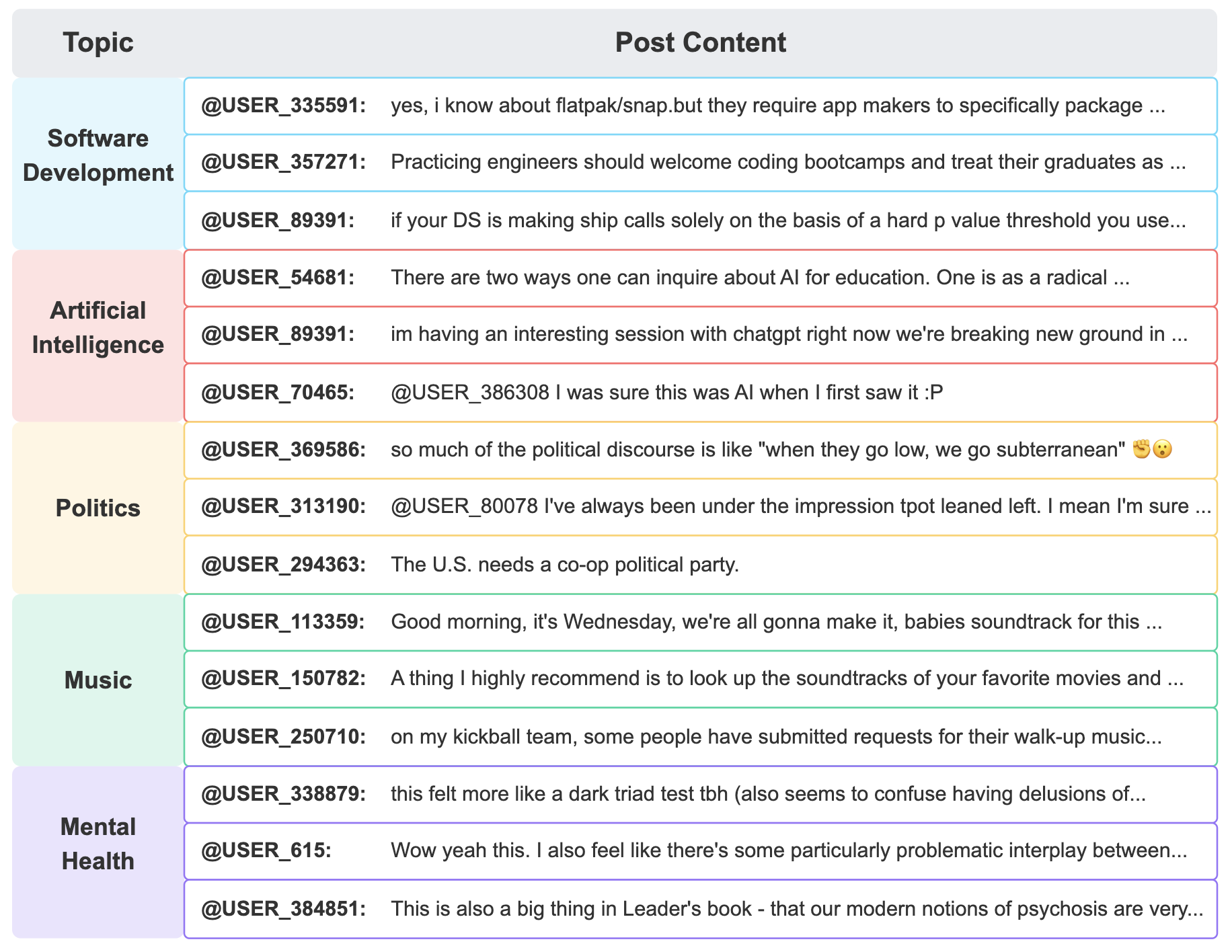}
\caption{3 sample posts from each of the top 5 most common topics in The Community Archive.}
\label{fig:samples}
\end{figure*}

\begin{figure*}[t]
    \centering
    \includegraphics[width=\linewidth]{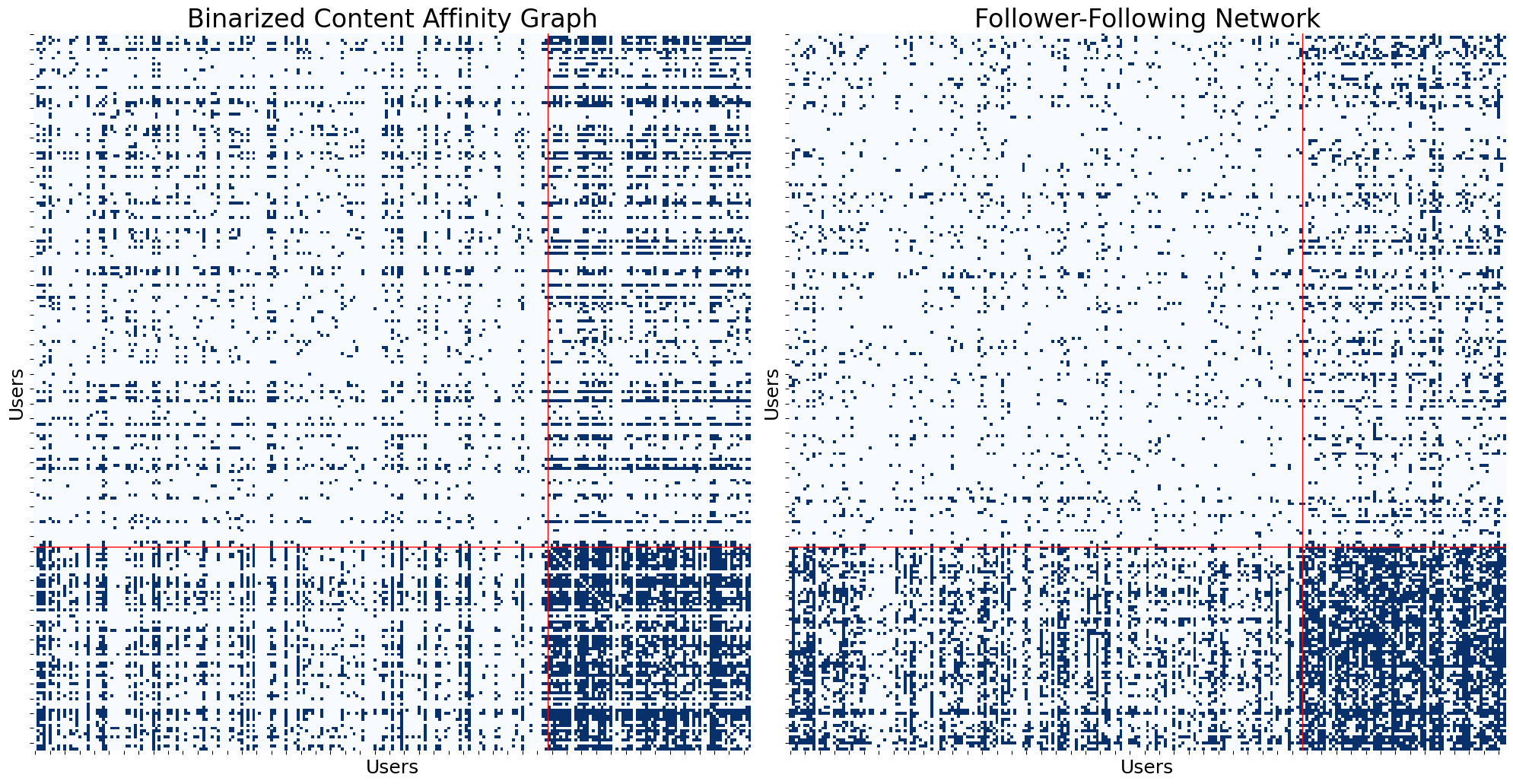}
        \caption{Comparison of the binarized content affinity graph (left) and the follower-following network (right) for users in The Community Archive. Users are sorted by cluster assignment from k-means clustering (k=2) applied to the follower-following adjacency matrix. The red lines denote the boundary between the inferred core and periphery groups. Both graphs contain the same number of edges and exhibit visual core-periphery structure, but the follower-following network displays a more pronounced core-periphery structure.}
    \label{fig:twoblock}
\end{figure*}

\subsection{Dataset Statistics}
\label{sec:stats}

The Community Archive contains 5,374,594 total posts across 244 unique accounts. 
The posts in our dataset are temporally concentrated, with 71.1\% of posts being made between January 1 2020 and June 1 2024.
We conjecture that the uptick in posting activity around January 2020 is a result of the COVID-19 Pandemic, and the decline in post activity after June 2024 is a result of decreasing archive upload activity.
Figure \ref{fig:twitstats} visualizes a histogram of post dates and archive upload activity.
A clear correspondence between decreasing archive upload activity and decreasing post availability can be seen.

The distribution of posts by author in our dataset is highly skewed, with the top 20\% of authors by post count making 64.21\% of the posts.
This skewness is consistent with several previous studies of X (Twitter) author activity \citep{wojcik2019sizing, pfeffer2023justdaytwittercomplete}.
We visualize these trends in Figure \ref{fig:twitstats}

To gain an understanding of the semantic composition of our dataset we followed the topic modeling methodology used in RedPajama \citep{weber2024redpajamaopendatasettraining} and computed topics using Nomic Atlas \citep{nomic2023atlas}. 
We post processed the topics computed in Nomic Atlas by merging duplicates, resulting in 598 topics and one additional "Miscellaneous" topic.
The number of posts in each of the 30 most frequent topics is visualized in Figure \ref{fig:topics}, and samples from each of the 5 most frequent topics are visualized in \ref{fig:samples}.
We release the computed topics for each tweet alongside our dataset to enable subsequent analysis.

\section{Example Application: Content Based Homophily}
In this section, we demonstrate the utility of our dataset by performing a simple analysis of content-based homophily in our network.
We begin by investigating aggregate content-based homophily, where we compare the temporally aggregated topic distribution between different groups in our network.
We then investigate temporal content-based homophily, where we compare the adoption rates of lexical innovations across different groups in our network.
This analysis is not meant to be extensive, but rather, is meant as an illustration of how The Community Archive can be used.
We release the code required to reproduce our analysis, which was originally run on an AWS EC2 c5.12xlarge machine.

\subsection{Aggregate Content-Based Homophily}
\label{sec:aggcohomo}
The Community Archive contains both post content and follower-following information.
This enables us to test the network for content-based homophily \citep{mcpherson2001homophily}, or the tendency of users that talk about similar topics to be connected in the graph.

We first generate a graph based on the content each user posts about. 
To construct this graph, we compute the distribution of the entirety of each user's content across each of the precomputed topics in The Community Archive.
This results in a time-aggregated distribution over topics for each user.
Next, we compute the Bhattacharyya coefficient between each pair of users content distributions.
This results in a weighted graph where every node is a user and edge weights correspond to similarity in the topics users post about.
Finally, we binarize this graph by retaining the top $9939$ edges by weight.
This ensures that the content graph and the follower-following graph have the same number of edges. 
The final binarized content affinity graph is shown in Figure \ref{fig:twoblock} (left).
The binarized content affinity graph visually exhibits core-periphery structure, but to a lesser degree than the follower-following graph.

We are interested in testing whether the binarized content affinity graph comes from the same distribution as the follower-following network.
We test this hypothesis in a parametric setting by assuming our follower-following network has a core-periphery structure \citep{yanchenko2023coreperipherystructurenetworksstatistical}, estimating the core-periphery model parameters from the follower-following network, and performing a bootstrap hypothesis test.

Formally, let $G_1$ be a graph with a node for each of the $n$ users in The Community Archive, and an edge between nodes $i$ and $j$ if user $i$ follows user $j$.
Let $G_2$ be the binarized content affinity graph over the same set of users.
Let $\hat z \in \{0, 1\}^n$ be the estimated community membership of each node, and $\hat P \in [0, 1]^{2\times2}$ be the estimated block connection probabilities.
We are interested in testing:

\[
\begin{aligned}
H_0 : \quad &G_2 \sim \mathrm{SBM}(\hat{z}, \hat{P})\\ 
H_A : \quad &G_2 \not\sim \mathrm{SBM}(\hat{z}, \hat{P})\\ 
\end{aligned}
\]

Where $\mathrm{SBM}(z, P)$ is a Stochastic Block Model with node memberships $z$ and block connections $P$.

We estimate community membership $\hat z$ via k-means clustering on the rows of the follower-following adjacency matrix with $k=2$, and estimate the block connection probabilities $\hat P$ using maximum likelihood.
Figure \ref{fig:twoblock} (right) shows the follower-following adjacency matrix, with rows and columns packed by cluster.
Core-periphery structure is clearly visible.

\begin{figure*}[t]
    \centering
    \includegraphics[width=0.7\linewidth]{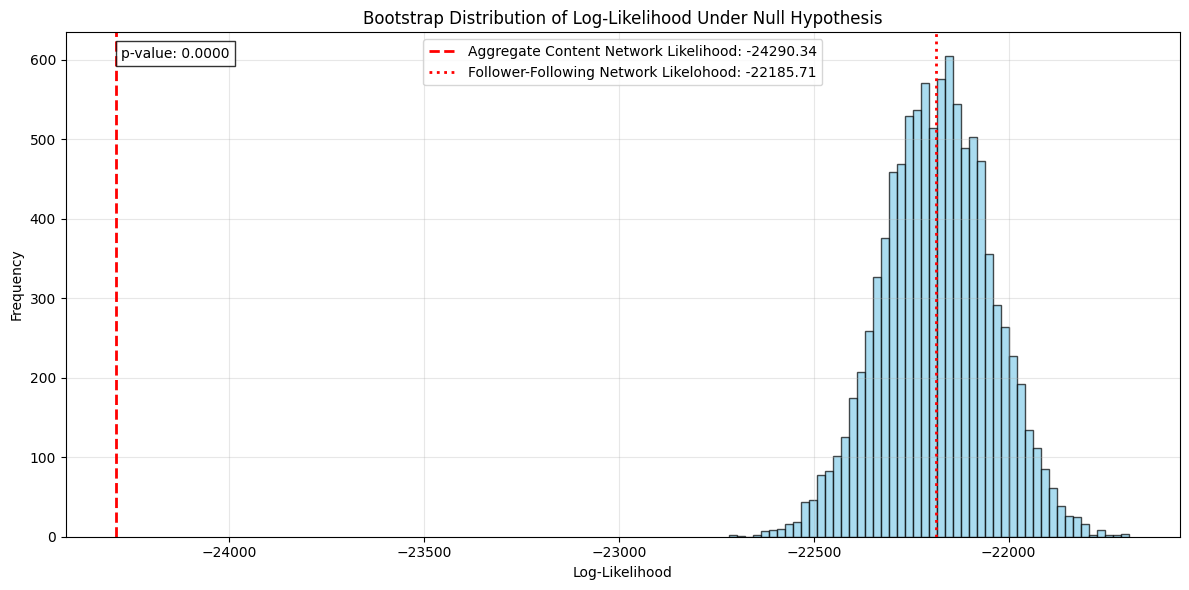}
    \caption{Bootstrap distribution of log-likelihoods under the null hypothesis. The red dashed line indicates the observed log-likelihood of the content affinity graph, which falls far outside of the left tail of the distribution. The red dotted line indicates the log-likelihood of the follower-following network, which falls directly in the center of the null distribution. With a p-value of 0, we reject the null hypothesis, providing evidence against aggregate content-based homophily under the assumed model.}
    \label{fig:hyptest}
\end{figure*}

To generate a null distribution, we perform a bootstrap \citep{efron1979bootstrap} using the core-periphery parameters estimated from our follower-following network.
We sample $10,000$ core-periphery networks from $\mathrm{SBM}(\hat z, \hat P)$, and compute each of their log-likelihoods.
We then compare the log-likelihood of the binarized content affinity graph to the distribution of log-likelihoods produced by the bootstrap.
The bootstrapped null distribution, as well as the test-statistic, are shown in Figure \ref{fig:hyptest}.

This procedure yields a p-value of $0$, as $G_2$ is highly unlikely under the SBM estimated from $G_1$.
Therefore, we reject the null-hypothesis, concluding that $G_2 \not\sim \mathrm{SBM}(\hat{z}, \hat{P})$.
We interpret this as a rejection of aggregate content-based homophily for this network.

We conjecture that one reason for this rejection is that members of the periphery talk about content originated by members of the core after a short time lag.
If this is the case, we would expect little difference in the content distributions of the core and periphery groups across a large enough time aggregation.

\begin{figure*}[t]
\centering
\begin{minipage}{\textwidth}
    \centering
    \begin{subfigure}{0.49\textwidth}
        \centering
        \includegraphics[width=\textwidth]{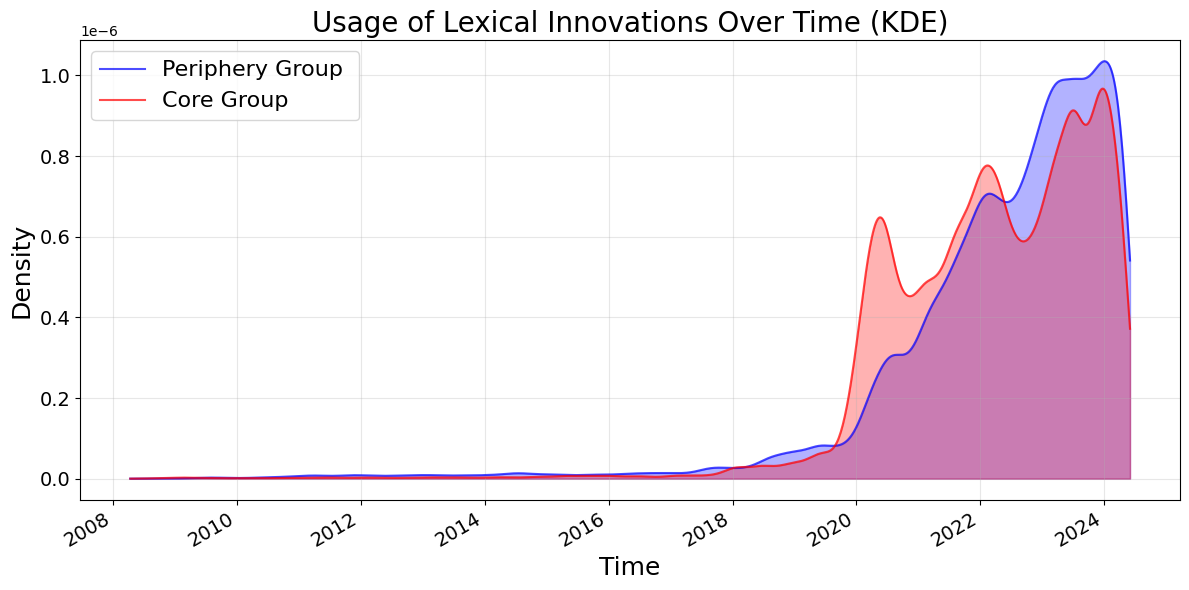}
    \end{subfigure}%
    \hspace{0.01\textwidth}
    \begin{subfigure}{0.49\textwidth}
        \centering
        \includegraphics[width=\textwidth]{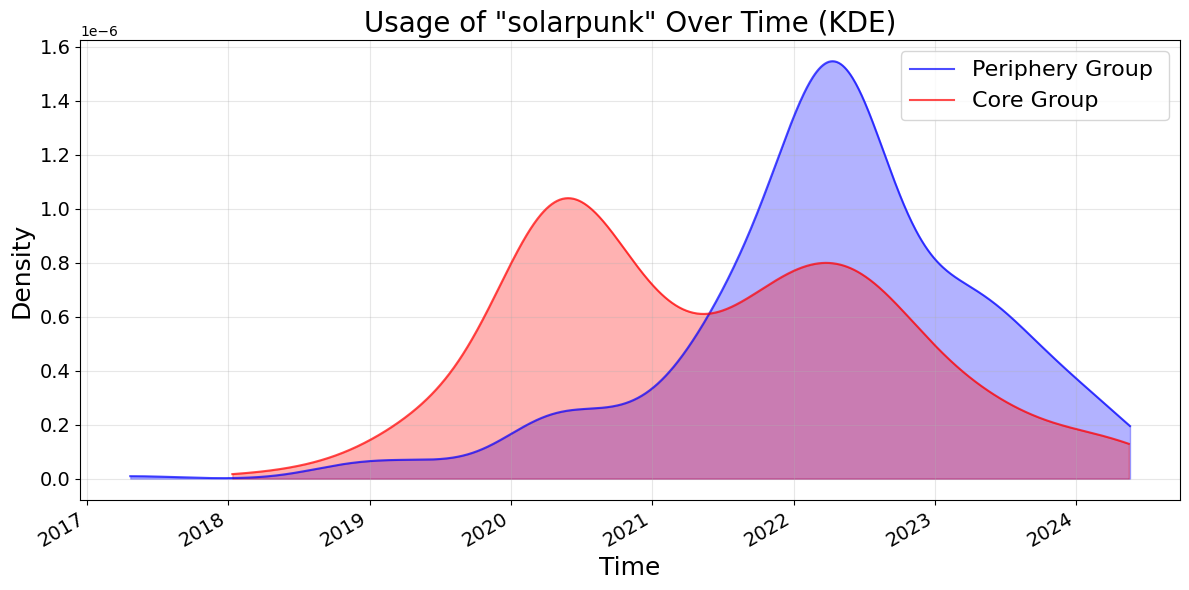}
    \end{subfigure}
\end{minipage}
\caption{\textbf{Left:} A KDE of timestamps when posts containing a lexical innovation were made, separated by group. The core group has a left shifted density relative to the periphery group, indicating they adopted lexical innovations earlier than the periphery group. \textbf{Right:} A KDE of timestamps when posts containing the the particular lexical innovation "solarpunk" were made. The core group adopts this lexical innovation before the periphery group.}
\label{fig:lexinnov}
\end{figure*}

\subsection{Temporal Content-Based Homophily}
\label{sec:narrative}
Given our rejection of aggregate content-based homophily in our network, we want to test the conjecture that members of the periphery talk about content originated by members of the core after a short time lag.
To investigate this, we track the spread of several lexical innovations across the network.
Given that this task is explicitly temporal, we utilize the temporal subset of our dataset.

To surface lexical innovations, we whitespace tokenize all of the posts in the dataset, and find tokens that did not occur before 2018, but occurred frequently after 2018.
We then focus on the top 100 of these tokens by number of unique users who have posted the token.
We exclude emojis, and analyze only text tokens.
We call the resulting 100 tokens a bag of lexical innovations.
A complete list of the bag of lexical innovations can be found in Appendix \ref{sec:lexbag}

Figure ~\ref{fig:lexinnov} (left) displays a Kernel Density Estimate (KDE) of timestamps when posts containing any term from our bag of lexical innovations were made, separated by core and periphery groups.
Visually, the distribution for the core group appears left-shifted relative to the periphery group, suggesting that members of the core may have adopted lexical innovations earlier.

To evaluate this statistically, we apply a one-sided Mann--Whitney U test~\cite{mann1947test}, comparing the distribution of adoption times between the two groups.
Let C denote the distribution of timestamps for lexical innovation usage by the core group, and 
P the corresponding distribution for the periphery group.
We test the null hypothesis that both distributions are equally likely to precede the other in adopting lexical innovations, against the alternative that the core group tends to adopt innovations earlier:
\begin{align*}
H_0 &: P(C < P) = 0.5 \\
H_A &: P(C < P) > 0.5
\end{align*}

This test yields a p-value of $6\mathrm{e}^{-13}$; therefore, we reject the null hypothesis.
We conclude that the core group adopted lexical innovations significantly earlier than the periphery group.
This finding supports the existence of temporal content-based homophily in our network, as lexical innovations tend to appear in the core before appearing in the periphery.

\section{Conclusion} 
We introduce The Community Archive, a novel dataset of over 5 million posts from the social media site X (formerly Twitter), collected entirely through an opt-in process that ensures both ethical and legal clarity.
Unlike prior large-scale social media datasets, which often lack either content, metadata, or user consent, The Community Archive offers all three, enabling studies into social network analysis, narrative tracking, and diachronic language modeling.

To demonstrate its utility, we conduct a sample analysis of content-based homophily in network.
Our analysis finds no evidence of aggregate content-based homophily under a core-periphery stochastic block model, but does reveal clear evidence of temporal content-based homophily; users in the core group tend to adopt lexical innovations before users in the periphery.
This finding underscores the value of temporally resolved, content-rich data for studying cultural transmission online.

By addressing the long-standing tension between research needs and ethical data collection, The Community Archive offers a new foundation for responsible, scalable, and transparent research.

\newpage
\newpage

\section{Limitations}
While The Community Archive provides an unprecedented ethically sourced corpus of X (Twitter) data, several limitations constrain its scope and generalizability.

First, the dataset suffers from a nontrivial sampling bias due to its opt-in collection methodology.
Users who chose to contribute their data are likely not representative of the broader Twitter population.
They may be more technically inclined, privacy aware, or embedded in certain ideological or professional communities.
This introduces structural and topical biases that may influence analyses, particularly those involving network topology or content diffusion.
In both scale and representativeness, The Community Archive is best understood as a "Zachary's Karate Club" for Twitter;
it is a dense, fully observable network that is nonetheless a small slice of the broader social graph.

Temporal resolution also presents a challenge.
Because archives were uploaded asynchronously, user activity in the dataset is only available up until the time of their export, with sharp declines in coverage thereafter.
While the temporal subset mitigates this issue for some analyses, it necessarily reduces the available data and does not fully eliminate time based distortions.

Moreover, the analyses we present focused on content based homophily and lexical innovation adoption are very simple.
Of particular note is the lack of a correction for the variance inherent in estimating the parameters of our core-periphery model in the hypothesis test of Section \ref{sec:aggcohomo}.
Much deeper analyses remain possible, and we hope future work will build on the foundation we laid in this paper.

Finally, while The Community Archive is the largest opt in corpus of its kind, it remains small compared to existing datasets collected without consent, some of which span hundreds of millions or even billions of posts.
As social media platforms increasingly restrict API access and raise legal barriers to research, it is imperative for the research community to develop broader, ethically sound solutions to the data access problem.
This need is particularly urgent given the accelerating entanglement of social platforms with political discourse, economic organization, and everyday life.

\section{Broader Impacts}
The Community Archive enables ethically grounded research on social media by providing a rare large-scale corpus of consented posts with both content and metadata.
By lowering legal and ethical barriers to analysis, it supports responsible work in areas like discourse modeling, network dynamics, and language change.
However, its limited scale and opt-in sampling bias may lead to skewed insights if used uncritically.
Broader adoption of ethical data collection practices remains essential as platforms increasingly restrict access.

\bibliography{anthology,custom}

\appendix

\section{Full Bag of Lexical Innovations}
\label{sec:lexbag}
covid, chatgpt, gpt, tpot, llm, llms, tiktok, vibecamp, pfp, lfg, psyop, dall, vibing, midjourney, nft, ngmi, poasting, retvrn, agentic, avoidant, slop, schizo, midwit, goated, clout, containment, jhanas, ineffable, bouba, maxxing, moots, doomer, wordcel, coronavirus, rizz, poasters, qiaochuyuan, disagreeable, poaster, nosilverv, simp, rlhf, playfulness, gpt4, iykyk, vaccinated, zoomers, yud, roon, qts, dalle, sbf, nickcammarata, modality, 200d, ashcroft, dissociated, hinge, speedrunning, zoomer, onlyfans, astral, adderall, tasshinfogleman, brainworms, bsky, airpods, attunement, walkable, gptbrooke, carbs, grindset, mimetic, mfs, reincarnation, algekalipso, solarpunk, gwern, enjoyer, wagmi, soaking, zoomed, frens, kamala, sashachapin, neuroses, deepfates, infohazards, hardship, burdens, uwu, dunks, blooming, others, imagines, psyops, epistemics, cognitively, rotators, justifications

\section{community-archive.org}
\begin{figure*}[h!]
    \centering
    \includegraphics[width=0.9\linewidth]{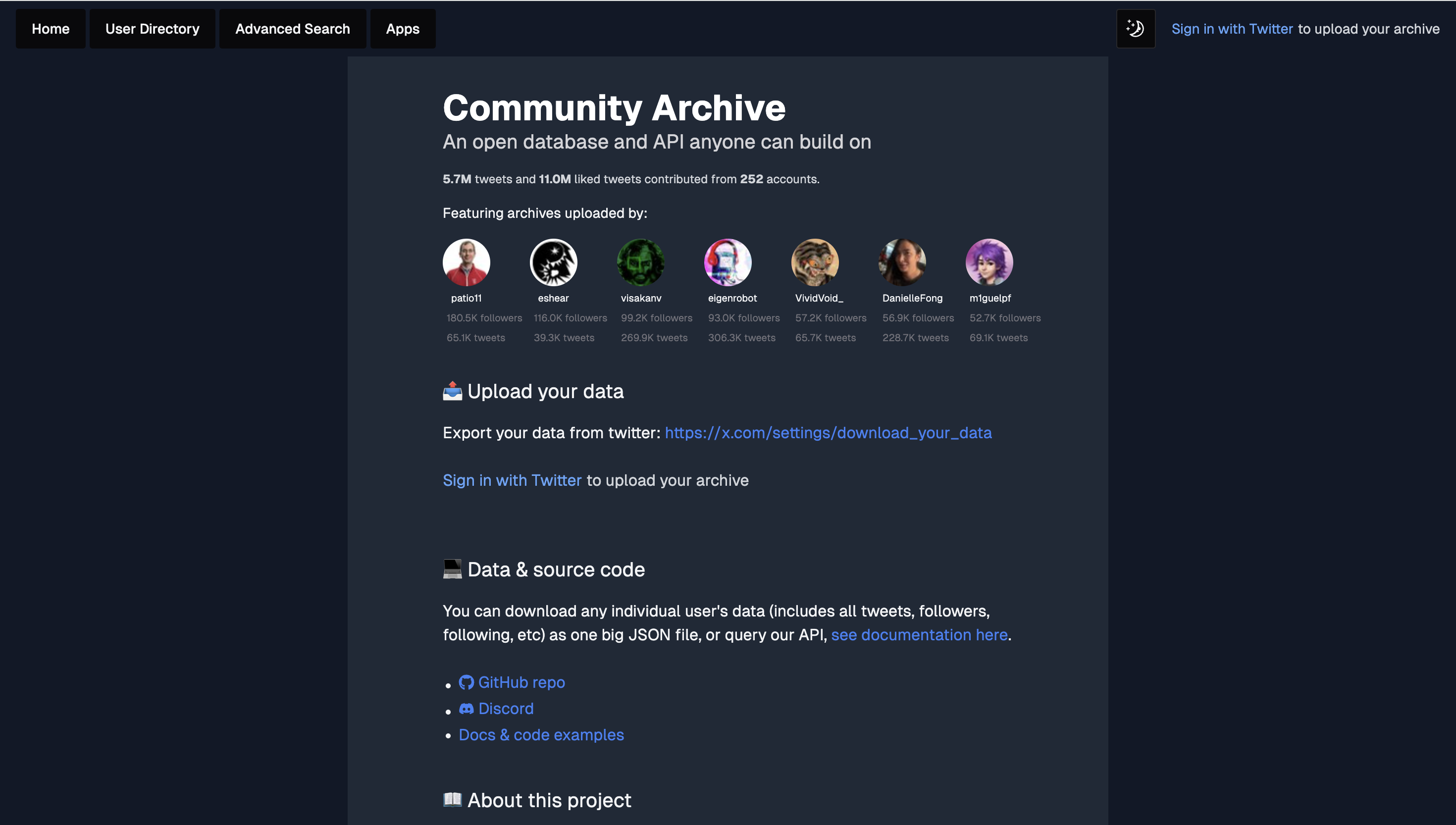}
    \caption{The data upload interface at community-archive.org. The open nature of the database and clear intent for future use by anyone are prominently displayed in the byline of the title.}
    \label{fig:ca-site}
\end{figure*}

\begin{figure*}[h!]
    \centering
    \includegraphics[width=0.9\linewidth]{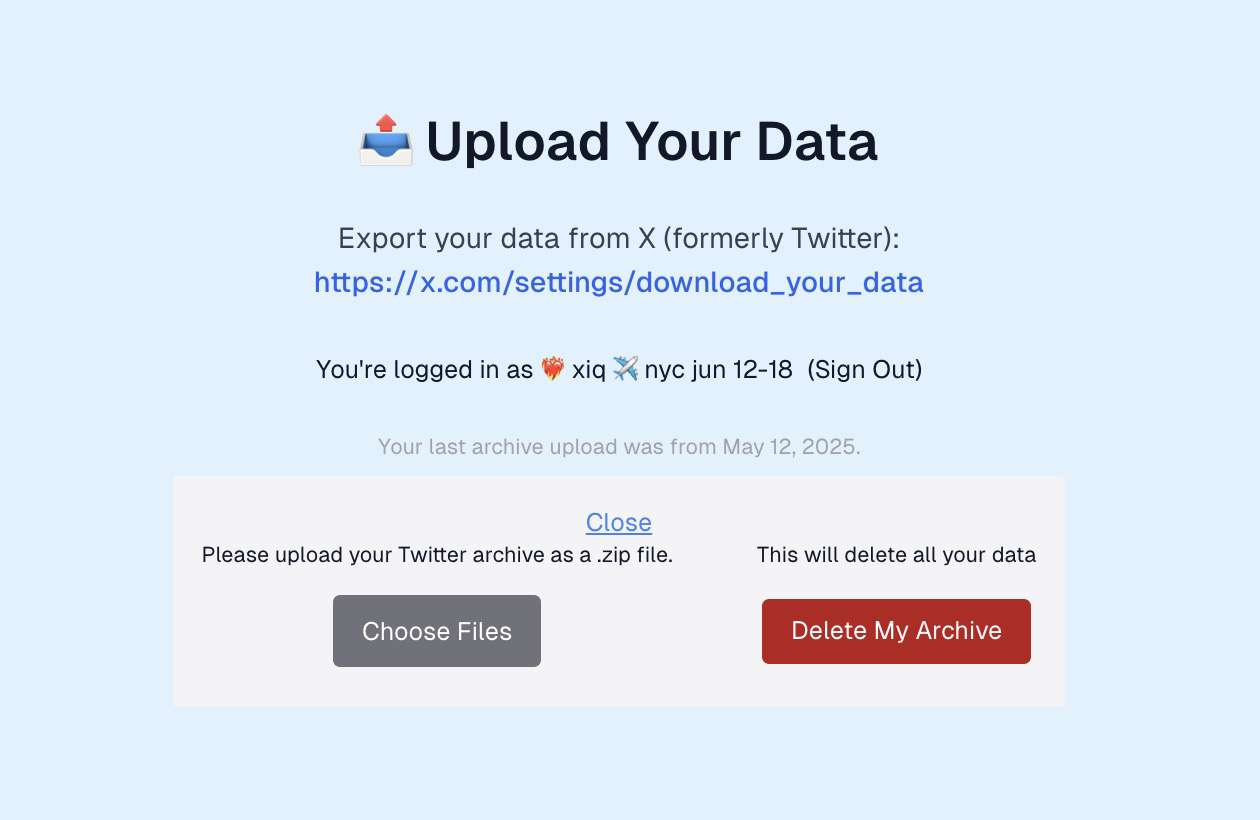}
    \caption{The delete archive option enables easy data removal. Screen is shown for a randomly selected sample user.}
    \label{fig:delete}
\end{figure*}

\section{Privacy Policy}
The following privacy policy is made visible to all community archive contributors:

\textbf{Community Archive Privacy Policy}

We are committed to preserving the public history of Twitter conversations while respecting your privacy.
This policy outlines how we handle your data and the options available to you.

\textbf{Data Collection and Usage}
The information we collect from your Twitter archive includes:
\begin{itemize}
    \item Profile information
    \item Tweets
    \item Likes
    \item Followers/following lists
\end{itemize}

We do not access or upload:
\begin{itemize}
    \item Direct messages
    \item Email addresses
    \item Deleted tweets
\end{itemize}

\textbf{Public Database and API}
By default, your uploaded archive becomes part of our public database and API.\textbf{ By uploading your data to the Community Archive, you are granting Epistemic Garden, Inc. the right to release your data under an MIT License.} This means:
\begin{itemize}
    \item Your tweets and likes will be visible to anyone.
    \item Researchers, developers, and other users can access and analyze this data.
    \item This data may be used for various purposes, including digital anthropology research and fine-tuning language models.
\end{itemize}

API docs \& instructions for downloading the data are in the GitHub repo.

\textbf{Important Considerations}
\begin{itemize}
    \item Data Accessibility: While your tweets are already public on Twitter, our platform makes them more easily accessible to a wider audience.
    \item Potential Risks: Be aware that malicious actors could potentially use this data to make inferences about your psychology or for targeted phishing attempts.
\end{itemize}

\textbf{Privacy Options}
We offer several options to give you more control over your data:
\begin{itemize}
    \item Exclude Likes: You can opt to leave out your likes when uploading your archive.
    \item Date Filtering: You can choose to make only a specific date range of your archive public.
    \item Future Controls: We plan to implement more granular permission settings in the future, such as allowing access to mutuals or specific individuals.
\end{itemize}

\end{document}